\documentstyle[aps,prb,epsf]{revtex} 
\begin {document} 
\draft 
 
\fnsymbol{footnote}
 
\title{\bf High--momentum dynamic structure function of liquid $^{\bf
3}$He--$^{\bf 4}$He mixtures: a microscopic approach}   
\author{F. Mazzanti \footnote{permanent address: Departament
d'Electr\`onica, Enginyeria La Salle, Pg. Bonanova, 8, Universitat Ramon
Llull, E-08022 Barcelona, Spain}}  
\address{Institut f\"ur Theoretische Physik, Johannes Kepler Universit\"at
Linz, A-4040 Linz, Austria}
\author{A. Polls}
\address{Departament d'Estructura i Constituents de la Mat\`eria, Diagonal
645, Universitat de Barcelona, E-08028 Barcelona, Spain}
\author{J. Boronat}
\address{ 
Departament de F\'{\i}sica i Enginyeria Nuclear, Campus Nord B4-B5,  
Universitat Polit\`ecnica de Catalunya. E-08034 Barcelona, Spain}  
\date{\today} 
 
\maketitle 
\begin{abstract} 
The high-momentum dynamic structure function of  liquid 
$^3$He-$^4$He mixtures
has been studied introducing final state effects. Corrections to the
impulse approximation have been included using a generalized
Gersch-Rodriguez theory that properly takes  into account the Fermi
statistics of $^3$He atoms. The microscopic inputs, as the momentum distributions
and the two-body density matrices, correspond to a variational
(fermi)-hypernetted chain calculation. The agreement with experimental data
obtained at $q=23.1$ \AA$^{-1}$ is not completely satisfactory, the
comparison being difficult due to inconsistencies present in the scattering
measurements. The significant differences between the experimental
determinations of the $^4$He condensate fraction and  the $^3$He kinetic
energy, and the theoretical results, still remain unsolved.
\end{abstract} 
 

\section{INTRODUCTION}
Liquid $^3$He-$^4$He mixtures at low temperature have been of long-standing
interest from both experimental\cite{ebner,edwards} and
theoretical\cite{kro,boro1,boni} viewpoints. From the
theoretical side, isotopic $^3$He-$^4$He mixtures manifest fascinating properties
intrinsically related to their quantum nature. The different quantum
statistics of $^4$He (boson) and $^3$He (fermion) appear reflected in the
macroscopic properties of the mixture as its very own existence in the 
zero-temperature limit. One of the most relevant features that the $^3$He-$^4$He
mixture shows is the interplay between both statistics driven by the
correlations. Signatures of
that are, on the one hand,  the influence of $^3$He on the condensate fraction 
($n_0$) and the
superfluid fraction ($\rho_s/\, \rho$) of $^4$He, and on the other, the
change in the momentum distribution of $^3$He atoms due to the correlations
with $^4$He. Both theory and experiment show that the $^4$He superfluid
fraction decreases with the $^3$He concentration ($x$) in the
mixture\cite{boni2,rhos}
whereas the condensate fraction $n_0$ moves on the opposite direction
showing an enhancement with $x$.\cite{wang} Concerning the $^3$He momentum
distribution in the mixture, microscopic calculations\cite{boro2,moro} point to a sizeable decrease in the
values of $n(k=0)$ and $Z=n(k_F^+)-n(k_F^-)$ with respect to pure $^3$He, with
a subsequent population at high $k$. The long tail of $n(k)$ gives rise to a
$^3$He kinetic energy which is appreciably larger than in pure $^3$He.

Experimental information on the momentum distribution $n(k)$ can be drawn
from deep inelastic neutron scattering (DINS),\cite{book1,book2} as first proposed by Hohenberg and
Platzman.\cite{hohen} It is nowadays well established that at high momentum transfer $q$ the
scattering is completely incoherent and accurately described by the impulse
approximation (IA). Assuming IA,
the momentum distribution can be directly extracted from experimental data.
However, this procedure is not  straightforward because of the
unavoidable instrumental resolution effects (IRE) and the non-negligible
final-state effects (FSE). The FSE are corrections to  IA that take into
account correlations between the struck atom and the medium which are
completely neglected in the IA. At the typical values of the momentum
transfer used in DINS on helium ($q \sim 20$ \AA$^{-1}$), both IRE and FSE
broaden significantly the IA prediction hindering a neat determination of
$n(k)$. The dominant contributions to the FSE are well accounted by the
different theoretical methods\cite{gers1,silver,carra} used in their study with an overall agreement
for $q \gtrsim 16$ \AA$^{-1}$.\cite{mazz1,rinat,fabrocini} Using the theoretical prediction for the FSE
and the IRE associated to the precision of the measurements, DINS in
superfluid $^4$He points to a condensate fraction $n_0=9.2 \pm 1.1$ and a
single-particle kinetic energy $T/N= 14.5 \pm 0.5$ K.\cite{azuaht} Both values are in a
nice agreement with the theoretical values obtained with Green's function
Monte Carlo (GFMC),\cite{kalos} diffusion Monte Carlo (DMC)\cite{boro3} and path integral Monte
Carlo (PIMC)\cite{cepe} methods.

Normal liquid $^3$He has also been studied by DINS.\cite{azuah} This system is more
involved from a technical point of view due to the large neutron absorption
cross section of $^3$He atoms which significantly reduces the
signal-to-noise ratio of the data. Recent measurements of $S(q,\omega)$ at
high $q$ point to a single-particle kinetic energy of $10 \pm 2$ K,\cite{azuaht,azuah}
a value that is clearly larger than a previous DINS determination 
($8.1 \pm 1.7$ K).\cite{sokol} A recent theoretical determination of
$T/N$ using DMC predicts\cite{casu} a value $12.24 \pm 0.03$ K in close agreement with
other microscopic calculations.\cite{panof,manu} Therefore, theory and experiment have
become closer but the agreement is still not so satisfactory as in liquid
$^4$He. These discrepancies have been generally attributed to high-energy tails
in $S(q,\omega)$, that are masked by the background noise, or even to
inadequacies of the gaussian models used to extract the momentum
distribution.\cite{paula}

In recent years, there have been a few  experimental studies of liquid 
$^3$He-$^4$He mixtures using DINS.\cite{wang,stirli} The response of the
mixture has been measured at two momenta, $q=23.1$ \AA$^{-1}$ and $q=110$ 
\AA$^{-1}$, and  different $^3$He concentrations. By using the
methodology employed in the analysis of the response of pure phases, results for the $^4$He condensate fraction
and the kinetic energies of the two species are extracted as a function of
$x$. This analysis points to a surprising result of
$n_0=0.18$,\cite{wang} a factor
two larger than in pure $^4$He. Concerning the kinetic energies, a
remarkable difference between $^4$He and $^3$He appears. The $^4$He kinetic
energy decreases linearly with $x$ whereas $^3$He atoms show a
$x$-independent kinetic energy that is the same of the pure $^3$He
phase.\cite{wang,stirli}
Except for the $^4$He kinetic energy, those experimental
measurements yield values that are sizeably different from the available theoretical
calculations. Microscopic approaches to the mixture using both
variational hypernetted-chain theory\cite{boro2} (HNC) and diffusion Monte 
Carlo\cite{moro} point
to a much smaller enhancement of $n_0$ in the mixture, and a $^3$He kinetic
energy much larger at small $x$ and that decrease with $x$ down to the pure
$^3$He result.

The theoretical estimation of the FSE in the scattering is of fundamental
interest, and also unavoidable in the analysis of the experimental data. In a
previous work,\cite{mazz1} we recovered the Gersch-Rodriguez (GR)
formalism\cite{gers1} for liquid
$^4$He, and proved that using  accurate approximations for the two-body
density matrix the FSE correcting function is very close to the predictions
of other FSE theories.\cite{silver,carra} The generalization of this theory to a Fermi system
as liquid $^3$He is not straightforward. The convolutive scheme developed
for a boson fluid is now impeded by the zeros present in the Fermi one-body density
matrix. Recently, it has been proposed an approximate GR-FSE theory that
incorporates the leading exchange contributions without the aforementioned
problems.\cite{mazz2} That theory is expected to capture the essential contributions
of Fermi statistics, and thereby to be accurate enough to generate the FSE
correcting function in dilute $^3$He-$^4$He liquid mixtures. The aim of the
present work is to provide microscopic results on the FSE effects in 
$^3$He-$^4$He mixtures. The inclusion of FSE on top of the impulse
approximation allows for a reliable prediction on the dynamic structure
function at high momentum transfer that can be compared with scattering data.

In the next section, the FSE formalism for the mixture is presented. Sect.
III is devoted to the lowest energy-weighted sum rules of the response and
the FSE correcting functions. The results and a comparison with available
experimental data are reported in Sect. IV. A brief summary and the main
conclusions comprise Sect. V.

\section{FSE IN FERMION-BOSON $^{\bf 3}$He--$^{\bf 4}$He LIQUID MIXTURES}
The dynamic structure function of the mixture $S(q,\omega)$ is completely
incoherent if the momentum transfer is high enough. The incoherent total
response can be split up in terms of the partial contributions
 $S^{(\alpha)}(q,\omega)$
\begin{equation}
S(q,\omega) = \sigma_4 (1-x) \, S^{(4)}(q,\omega) + \sigma_3 x \, 
S^{(3)}(q,\omega) \  ,
\label{sinc}
\end{equation}
$x=N_3/\,N$ being the $^3$He concentration, and $\sigma_4$, $\sigma_3$ the
cross sections of the individual scattering processes ($\sigma_3=5.61$
barn, $\sigma_4=1.34$ barn). Notice that in this regime the cross
term $S^{(3,4)}(q,\omega)$ does not appear because it is fully coherent, and
the incoherent density- and spin-dependent Fermi responses are identical (
$\sigma_3= \sigma_{3,d} + \sigma_{3,I}$ with $\sigma_{3,d}=4.42$ barn and 
$\sigma_{3,I}=1.19$ barn). In Eq. (\ref{sinc}), each single term can be
obtained as the Fourier transform of the corresponding density-density
correlation factor
\begin{equation}
S^{(\alpha)}(q,t) = \frac{1}{N_\alpha} \sum_{j=1}^{N_\alpha} \langle
e^{-i {\bf q} \cdot {\bf r}_j} e^{iHt} e^{i {\bf q} \cdot {\bf r}_j}
e^{-iHt}  \rangle \ ,  
\label{sden}
\end{equation}
where $\alpha=$3, 4 stands for $^3$He and $^4$He, respectively. In Eq.
(\ref{sden}), $H$ is the hamiltonian of the system ($\hbar=1$),
\begin{equation}
H=- \frac{1}{2m_4} \sum_{j=1}^{N_4} \nabla_j^2 - \frac{1}{2m_3}
\sum_{j=1}^{N_3} \nabla_j^2  + \frac{1}{2}\, \sum_{\alpha,\beta=3,4}
\sum_{i,j=1}^{N_3,N_4} V^{(\alpha,\beta)}(r_{ij})   \ ,
\label{hamil}
\end{equation}
with $V^{(\alpha,\beta)}(r_{ij})$ the pair-wise interatomic potentials, that
in an isotopic mixture, as the present one, are all identical.

The  translation operators  act on the hamiltonian H and transforms Eq.
(\ref{sden}) into
\begin{equation}
S^{(\alpha)}(q,t) = \frac{1}{N_\alpha} e^{i\omega_q^{(\alpha)} t}
\sum_{j=1}^{N_\alpha} \langle e^{i(H+L_j^{(\alpha)})t} e^{-iHt} \rangle \ ,
\label{sden2}
\end{equation}
with $\omega_q^{(\alpha)}=q^2/\,(2 m_\alpha)$, and $L_j^{(\alpha)}= {\bf
v}^{(\alpha)} \cdot {\bf p}_j$ being the projection of the momentum of particle
$j$ along the direction of the recoiling velocity ${\bf v}^{(\alpha)}= 
{\bf q}/ m_\alpha$. One can then define a new operator ${\cal C}^{(\alpha)}(t)$
\begin{equation}
{\cal C}^{(\alpha)}(t) \equiv e^{-iHt} e^{i(H+L_1^{(\alpha)})t}
e^{-iL_1^{(\alpha)}t} 
\label{calfa}
\end{equation}
that contains the FSE corrections to the IA. In terms of ${\cal
C}^\alpha(t)$, the
density-density correlation factor turns out to be
\begin{equation}
S^{(\alpha)}(q,t) = e^{i \omega_q^{(\alpha)} t} \langle {\cal
C}^{(\alpha)}(t) 
\,e^{i t L_1^{(\alpha)}} \rangle \ .
\label{sden3}
\end{equation}

In the  high-momentum transfer regime, in which we are interested in,
${\bf v}^{(\alpha)}$ is large while $t$ is short, in such a way
that their product $s={\bf v}^{(\alpha)} t$ is of order one. In terms of
this new variable $s$, Eqs. (\ref{calfa}) and (\ref{sden3}) become
\begin{equation}
S^{(\alpha)}(q,s) = e^{i s \omega_q^{(\alpha)} / v^{(\alpha)}} \langle 
{\cal C}^{(\alpha)}(s) 
\,e^{i s \hat{{\bf v}}^{(\alpha)}\cdot{\bf p}_1} \rangle \ ,
\label{sden4}
\end{equation}
\begin{equation}
{\cal C}^{(\alpha)} (s) = e^{-i s{\cal H}} 
e^{i s({\cal H}+\hat{\bf v}^{(\alpha)} \cdot {\bf p}_1)}
e^{-i s \hat{\bf v}^{(\alpha)} \cdot {\bf p_1} } \ ,
\label{calfa2}
\end{equation}
with a new hamiltonian ${\cal H} = H / v^{(\alpha)}$. The operators 
${\cal C}^{(\alpha)}(s)$ satisfy the differential equation
\begin{equation}
\frac{d}{d s} {\cal C}^{(\alpha)} (s)  = 
i [ \Lambda^\dagger(s) {\cal H} \Lambda(s) - {\cal H} ] \, 
{\cal C}^{(\alpha)}(s) 
\ ,
\label{diff}
\end{equation}
with
\begin{equation}
\Lambda^\dagger(s) \equiv e^{-i s {\cal H}} 
e^{i s({\cal H}+\hat{\bf v}^{(\alpha)}\cdot{\bf p}_1) } \ .
\label{lambda}
\end{equation}

The differential equation (\ref{diff}) may be solved by means of a cumulant expansion
in powers of $1/v^{(\alpha)}$.\cite{mazz2} In the high-$q$ limit, only the first terms
of the resulting series are expected to significantly contribute. In fact,
the IA is recovered when only the zero-order term is retained, 
\begin{equation}
S_0^{(\alpha)}(q,s) = e^{i \omega_q^{(\alpha)}/ v^{(\alpha)} }
\frac{1}{\rho_\alpha} \rho_1^{(\alpha)}(s) \ ,
\label{ia1}
\end{equation}
$\rho_1^{(\alpha)}(s)$ being the one-body density matrix. By including the
next-to-leading term, the leading corrections (FSE) to the IA are taken into
account,
\begin{equation}
S_1^{(\alpha)}(q,s) = e^{i s \omega_q^{(\alpha)}/ v^{(\alpha)} }
\left \langle e^{i \frac{1}{v^ {(\alpha)}} \int_0^s (H_0(s^\prime) - H) d
s^\prime } e^{i {\bf s} \cdot {\bf p}_1 } \right \rangle        \ ,
\label{fse1}
\end{equation}
with $H_0(s) = e^{i {\bf s} \cdot {\bf p}_1} H 
e^{-i {\bf s} \cdot {\bf p}_1} $. A Gersch-Rodriguez cumulant expansion
of Eq. (\ref{fse1}) for the $^4$He component in the mixture leads to a FSE
convolutive scheme
\begin{equation}
S_1^{(4)}(q,s) = S_{\rm IA}^{(4)}(q,s) \, R^{(4)}(q,s) \ ,
\label{fsehe4}
\end{equation}
with the IA response (\ref{ia1}), and
\begin{eqnarray}
R^{(4)}(q,s)  & = & 
\exp\Biggl[-\frac{1}{\rho_1^{(4)}(s)} \int d{\bf r} \, 
\rho_2^{(4,4)}({\bf r}, 0; {\bf r}+{\bf s}) 
\left[ 1 - \exp\left( \frac{i}{v^{(4)}} \int_0^s ds^\prime  
\Delta V({\bf r}, {\bf s}^\prime) \right)\right] 
\nonumber  \\ 
&  & 
-\frac{1}{\rho_1^{(4)}(s)} \int d{\bf r} \, 
\rho_2^{(4,3)}({\bf r}, 0; {\bf r}+{\bf s}) 
\left[ 1 - \exp\left( \frac{i}{v^{(4)}} \int_0^s ds^\prime
\Delta V({\bf r}, {\bf s}^\prime) \right)\right] \Biggr] \ .
\label{rcon4}
\end{eqnarray}
In the above equation, $\Delta V({\bf r}_{ij},{\bf r}^\prime) 
\equiv  V({\bf r}_{ij}+{\bf r}^\prime) - V(r_{ij})$. Apart from
$\rho_1^{(\alpha)}$, $R^{(4)}(q,s)$ is a function of the (4,4) and (4,3)
components of the semi-diagonal two-body density matrix
\begin{equation}
\rho_2^{(\alpha,\beta)}({\bf r}_1, {\bf r_2}; {\bf r}'_1, {\bf r}_2) =
N_\alpha \left( N_\beta - \delta_{\alpha \beta} \right) 
\frac{\int {\rm d}{\bf r}^{N-2}\, \Psi_0^*({\bf r}_1, {\bf r_2}, \ldots, 
{\bf r}_N) \Psi_0({\bf r}'_1, {\bf r}_2, \ldots, {\bf r}_N)}
{\int {\rm d}{\bf r}^N \mid \Psi({\bf r}_1, {\bf r}_2, \ldots, {\bf r}_N)
\mid^2 } \ .
\label{2body}
\end{equation}
 
The analysis of the $^3$He (fermion) component is much more involved. A
fully convolutive formalism is now forbidden because the zero-order
cumulant, which is proportional to the one-body density matrix, has an
infinite number of nodes. Nevertheless, it is plausible to assume that at
high $q$ the FSE are dominated by dynamical correlations, and that
statistical corrections to a purely FSE scheme can therefore be introduced
perturbatively. With this hypothesis, the $^3$He response can be split up in two
terms,\cite{mazz2}
\begin{equation}
S^{(3)}(q,s) \equiv S_{\rm B}^{(3)}(q,s) + \Delta S^{(3)}(q,s) \ ,
\label{split3}
\end{equation}
using the following identity for the $n$-body density matrix of the mixture
\begin{eqnarray}
\rho_N({\bf r}_1, {\bf r}_2, \ldots, {\bf r}_N; {\bf r}'_1) 
 & = & 
\rho_1^{(3)}({\bf r}_{11'})\,\left[ \frac{1}{\rho_1^{\rm B}(r_{11'})}\,
\rho_N^{\rm B}({\bf r}_1, {\bf r}_2, \ldots, {\bf r}_N; {\bf r}'_1) \right] 
\label{rhoN}  \\ 
& + &  \left[ 
\rho_N({\bf r}_1, {\bf r}_2, \ldots, {\bf r}_N; {\bf r}'_1) 
- \frac{\rho_1^{(3)}(r_{11'})}{\rho_1^{\rm B}(r_{11'})} 
\rho_N^{\rm B}({\bf r}_1, {\bf r}_2, \ldots, {\bf r}_N; {\bf r}'_1) 
\right]  \ .
\nonumber
\end{eqnarray}
The superscript B stands for a boson approximation, i.e., a fictitious 
boson-boson $^3$He-$^4$He mixture. In that factorization (\ref{rhoN}), the
first term allows for a description of the $^3$He response in which the IA
is the exact one while the FSE are introduced in a boson-boson approximation.
Statistical corrections to the FSE are all contained in the second term.

In Eq. (\ref{split3}), $S_{\rm B}^{(3)}(q,s)$ is the main part of the
response and can be written as a convolution product
\begin{equation}
S_{\rm B}^{(3)}(q,s) = S_{\rm IA}^{(3)}(q,s) \, R^{(3)}(q,s) \ ,
\label{convol3}
\end{equation}
with $S_{\rm IA}^{(3)}(q,s)  =   
e^{i s \omega_q^{(3)}/ v^{(3)}} \rho_1^{(3)}(s) / \rho_3$ the impulse
approximation, and
\begin{eqnarray}
R^{(3)}(q,s)  & = & 
\exp\Biggl[ 
-\frac{1}{\rho_1^{\rm B}(s)} \int d{\bf r}\, 
\rho_2^{(3,3){\rm B}}({\bf r}, 0; {\bf r}+{\bf s})
\left[ 1 - \exp\left( \frac{i}{v^{(3)}}\int_0^s ds^\prime
\Delta V({\bf r}, {\bf s}^\prime) \right) \right]
\nonumber  \\ 
& &  -\frac{1}{\rho_1^{\rm B}(s)} \int d{\bf r} \,
\rho_2^{(3,4){\rm B}}({\bf r}, 0; {\bf r}+{\bf s})
\left[ 1 - \exp\left( \frac{i}{v^{(3)}}\int_0^s ds^\prime
\Delta V({\bf r}, {\bf s}^\prime) \right) \right] \Biggr] 
\label{fse3}
\end{eqnarray}
the boson-like FSE correcting function.

The additive correction $\Delta S^{(3)}(q,s)$ in Eq. (\ref{split3}) 
takes into account the statistical exchange
contributions in the FSE and is expected to be small. 
Actually, it is a function of
\begin{equation}
\Delta \rho_2^{(3,\alpha)}({\bf r}_1,{\bf r}_2;{\bf r}_1^\prime) =
\rho_2^{(3,\alpha)}({\bf r}_1,{\bf r}_2;{\bf r}_1^\prime) -
\frac{\rho_1^{(3)}(r_{11^\prime})}{\rho_1^{\rm B}(r_{11^\prime})} \,
\rho_2^{(3,\alpha){\rm B}}({\bf r}_1,{\bf r}_2;{\bf r}_1^\prime) \ ,
\label{drho3}
\end{equation}
according to the decomposition (\ref{rhoN}).
The variational framework
of the (fermi)-hypernetted chain equations (F)HNC that is used in this work
to calculate the one- and two-body density matrices, provides a
diagrammatic expansion to estimate $ \Delta \rho_2^{(3,\alpha)}$. Following
the diagrammatic rules of the FHNC/HNC formalism, $ \Delta \rho_2^{(3,\alpha)}$
may be written as the sum of two terms:
\begin{equation}
\Delta \rho_2^{(3,\alpha)}({\bf r}_1,{\bf r}_2;{\bf r}_1^\prime) =
\rho_\alpha \rho_1^{(3)}(r_{11^\prime}) G^{(3,\alpha)}
({\bf r}_1,{\bf r}_2;{\bf r}_1^\prime) -
\rho_\alpha \rho_{1{\rm D}}(r_{11^\prime}) F^{(3,\alpha)}
({\bf r}_1,{\bf r}_2;{\bf r}_1^\prime) \ .
\label{drho3-2}
\end{equation}
$\rho_1^{(3)}(r)$ is the one-body density matrix and $\rho_{1{\rm D}}(r)$ is an
auxiliary function, which factorizes in $\rho_1^{(3)}(r)$, and that sums up
all the diagrams contributing to  $\rho_1^{(3)}(r)$ except those where the external
points 1 and 1$^\prime$ are statistically linked.\cite{mazphd} $F^{(3,\alpha)}$ and 
$G^{(3,\alpha)}$ in Eq. (\ref{drho3-2}) sum up diagrams with the external
vertices (1,1$^\prime$,2) with and without statistical lines attached to 1
and 1$^\prime$, respectively. With this prescription for 
$\Delta \rho_2^{(3,\alpha)}$, the additive term $\Delta S^{(3)}(q,s)$
becomes finally
\begin{eqnarray}
\Delta S^{(3)}(q,s)  & = & 
e^{i s \omega_q^{(3)}/v^{(3)}} \frac{1}{\rho_3}\,\rho_{1{\rm D}}(s)
\label{delta3l}  \\ 
& \times &  \Biggl[\,
\exp\Biggl[ -\frac{1}{\rho_{1{\rm D}} (s)} \int d{\bf r}\,
\Delta \rho_2^{(3,3)}({\bf r}, 0; {\bf r}+{\bf s}) 
\left[ 1-\exp\left( \frac{i}{v^{(3)}} \int_0^s ds^\prime
\Delta V({\bf r}, {\bf s}^\prime) \right) \right] 
\nonumber  \\ 
&  & -
\frac{1}{\rho_{1{\rm D}} (s)} \int d{\bf r}\,
\Delta \rho_2^{(3,4)}({\bf r}, 0; {\bf r}+{\bf s}) 
\left[ 1-\exp\left( \frac{i}{v^{(3)}} \int_0^s ds^\prime
\Delta V({\bf r}, {\bf s}^\prime) \right) \right] - 1 \Biggr] \ .
\nonumber
\end{eqnarray}

Equations (\ref{rcon4}), (\ref{fse3}), and (\ref{delta3l}) are the final
results of the present theory for the FSE in $^3$He-$^4$He mixtures. They
constitute the generalization of the Gersch-Rodriguez formalism to a
mixture with special emphasis in the difficulties arising from Fermi
statistics. Apart from the interatomic potential, very well-known in helium,
the microscopic inputs that are required are the one- and two-body density
matrices, both in the boson-boson and the fermion-boson cases. 

To conclude this section, we define the Compton
profiles of each component in the mixture. Contrarily to what happens in 
a pure phase, the total response of the mixture can  not be written in terms of a single scaling
variable $Y$. Each individual profile is naturally given in its own scaling
variable $Y_\alpha = m_\alpha \omega / q - q/2$. Thus,
\begin{equation}
J^{(\alpha)}(q,Y_\alpha) = \frac{1}{2 \pi} \, \int_{-\infty}^{\infty} d s
\, e^{- i Y_\alpha s} S^{(\alpha)}(q,s)   \ ,
\label{jalfa}
\end{equation}
which after introducing the explicit expressions for $S^{(\alpha)}(q,s)$ 
becomes 
\begin{equation}
J^{(\alpha)}(q,Y_\alpha) = \int_{-\infty}^{\infty} d Y_\alpha \,
J^{(\alpha)}(Y_\alpha) R^{(\alpha)}(q,Y_\alpha) + 
\Delta J^{(3)}(q,Y_\alpha) \delta_{\alpha 3} \ .
\label{jalfat}
\end{equation} 
In this equation, $\Delta J^{(3)}$ derives from $\Delta S^{(3)}$ and the IA
responses $J^{(\alpha)}(Y_\alpha)$ are directly related to the momentum
distributions $n^{(\alpha)}(k)$
\begin{equation}
J^{(\alpha)}(Y_\alpha) = n_0 \delta(Y_4) \delta_{\alpha 4} +
\frac{\nu_\alpha}{4 \pi^2 \rho_\alpha} \int_{|Y_\alpha|}^{\infty} dp \, p
n^{(\alpha)}(p) \ ,
\label{jia}
\end{equation} 
$n_0$ being the $^4$He condensate fraction, and $\nu_\alpha$ the spin
degeneracy of each component ($\nu_3=2$, $\nu_4=1$). Notice that the first
term in Eq. (\ref{jalfat}) contains the explicit contribution
$ n_0 R^{(4)}(q,Y_\alpha) \delta_{\alpha 4}$ arising from the condensate.

\section{ENERGY--WEIGHTED SUM RULES AT HIGH MOMENTUM TRANSFER}
Energy-weighted sum rules provide an useful tool to analyze the properties
of $S(q,\omega)$. In spite of the fact
that the knowledge of a small set of energy moments usually is not enough to
completely characterize the response, the method has proved its usefulness
in the analysis of scattering on quantum fluids.\cite{stringa,dalfo} Moreover, from a
theoretical viewpoint the comparison between the sum rules derived from an
approximate theory and the exact ones shed light on the accuracy of that
approach. In the high-$q$ limit, the response is fully incoherent and
therefore we discuss only the incoherent sum rules
\begin{equation}
m_n^{(\alpha)}(q)  =  \int_{- \infty}^{\infty} d \omega \, \omega^n S_{\rm
inc}^{(\alpha)}(q,\omega)  = \frac{1}{i^n} \frac{d^n}{dt^n} S_{\rm
inc}^{(\alpha)}(q,t) \, |_{t=0}  \ .
\label{suminc}
\end{equation}

Considering
\begin{equation}
S_{\rm inc}^{(\alpha)}(q,t) = \langle e^{-i {\bf q}\cdot {\bf
r}_1^{(\alpha)}} e^{i Ht} e^{i {\bf q}\cdot {\bf r}_1^{(\alpha)}}
e^{-i Ht}  \rangle  \ ,
\label{sinc2}
\end{equation}
and applying to the three rightmost operators in (\ref{sinc2}) the
Baker-Campbell-Hausdorff formula one arrives to the following expansion in terms of
$it$:
\begin{eqnarray}
S_{\rm inc}^{(\alpha)}(q,t) = 1 & + & it \langle e^{-i {\bf q}\cdot {\bf
r}_1^{(\alpha)}} \, [H,e^{i {\bf q}\cdot {\bf r}_1^{(\alpha)}} ] \rangle
\nonumber \\
 & + & \frac{1}{2!} (it)^2 \langle e^{-i {\bf q}\cdot {\bf
r}_1^{(\alpha)}} [ H, [H,e^{i {\bf q}\cdot {\bf r}_1^{(\alpha)}} ] ]
\rangle  \nonumber \\
 & + &  \frac{1}{3!} (it)^3 \langle e^{-i {\bf q}\cdot {\bf
 r}_1^{(\alpha)}} [ H, [H, [H, e^{i {\bf q}\cdot {\bf r}_1^{(\alpha)}}]]]
 \rangle + \cdots
\label{sinc3}
\end{eqnarray}
From Eqs. (\ref{suminc}) and (\ref{sinc3}), one easily identifies the lowest-order
sum rules:
\begin{eqnarray}
m_{0,{\rm inc}}^{(\alpha)} (q) &  =  & 1    
\label{m0inc} \\
m_{1,{\rm inc}}^{(\alpha)} (q) &  =  & \langle e^{-i {\bf q}\cdot {\bf
r}_1^{(\alpha)}} \, [H,e^{i {\bf q}\cdot {\bf r}_1^{(\alpha)}} ] \rangle =
\frac{q^2}{2m_\alpha}    
\label{m1inc} \\
m_{2,{\rm inc}}^{(\alpha)} (q) & = & \langle e^{-i {\bf q}\cdot {\bf
r}_1^{(\alpha)}} [ H, [H,e^{i {\bf q}\cdot {\bf r}_1^{(\alpha)}} ] ]
\rangle = \left( \frac{q^2}{2m_\alpha} \right)^2 + 
\frac{4}{3} \frac{q^2}{2m_\alpha}
t_\alpha
\label{m2inc} \\
m_{3,{\rm inc}}^{(\alpha)} (q) & = & \langle e^{-i {\bf q}\cdot {\bf
 r}_1^{(\alpha)}} [ H, [H, [H, e^{i {\bf q}\cdot {\bf r}_1^{(\alpha)}}]]]
 \rangle  =   
 \label{m3inc} \\
& = &  \left ( \frac{q^2}{2m_\alpha} \right )^3 + 
4 \left ( \frac{q^2}{2m_\alpha} \right )^2 \, t_\alpha +
\frac{1}{2m_\alpha} \rho \int d{\bf r} \, g^{(\alpha,\alpha)}(r) ({\bf q}
\cdot \nabla )^2 V(r)   \nonumber 
\end{eqnarray}
All four moments can be readily calculated from the interatomic pair
potential $V(r)$, the kinetic energies per particle $t_\alpha$, and the
two-body radial distribution function between pairs of atoms of the same
kind $g^{(\alpha,\alpha)}(r)$. $m_{1,{\rm inc}}^{(\alpha)} (q)$ is
identical to the total $m_{1}^{(\alpha)} (q)$, also known as the f-sum
rule, whereas the other three coincide with the leading contribution to the
total sum rules $m_{n}^{(\alpha)} (q)$ at high $q$.

In the limit $q \rightarrow \infty$  the IA is expected to be the
dominant term. This feature may be analyzed using the sum-rules
methodology. Starting from the IA response
\begin{equation}
S_{\rm IA}^{(\alpha)} (q,\omega) = \frac{\nu_\alpha}{(2 \pi)^3 \rho_\alpha}
\, \int d{\bf k} \, n^{(\alpha)}(k) \ \delta \left( \frac{({\bf q} + {\bf
k})^2}{2m_\alpha} - \frac{k^2}{2m_\alpha} - \omega \right )  \ ,
\label{siaalpha}
\end{equation}
one can calculate the first energy moments from basic properties of the
momentum distributions. The results are:
\begin{eqnarray}
m_{0,{\rm IA}}^{(\alpha)} (q) &  =  & 1    
\label{m0IA} \\
m_{1,{\rm IA}}^{(\alpha)} (q) &  =  & 
\frac{q^2}{2m_\alpha}    
\label{m1IA} \\
m_{2,{\rm IA}}^{(\alpha)} (q) & = & \left ( \frac{q^2}{2m_\alpha} \right
)^2 + \frac{4}{3} \frac{q^2}{2m_\alpha}
t_\alpha
\label{m2IA} \\
m_{3,{\rm IA}}^{(\alpha)} (q) & = &  
 \left ( \frac{q^2}{2m_\alpha} \right )^3 + 
4 \left ( \frac{q^2}{2m_\alpha} \right )^2 \, t_\alpha  
\label{m3IA}
\end{eqnarray}
When the IA sum rules are compared with the incoherent results
(\ref{m0inc},\ref{m1inc},\ref{m2inc},\ref{m3inc}), one realizes that the first three moments
are exhausted by IA. The leading order terms in $q$ in the $m_3$ sum rule
are also reproduced by the IA but the term with $g^{(\alpha,\alpha)}(r)$ is
not recovered.

The variable that  naturally emerges in the $1/q$ expansion of the response
of the mixture is the West scaling variable $Y_\alpha$. It is therefore
also useful to consider the
$Y_\alpha$-weighted sum rules of $J^{(\alpha)}(q,Y_\alpha)$
\begin{equation}
M_n^{(\alpha)}(q)  =  \int_{- \infty}^{\infty} d Y_\alpha \, Y_\alpha^n
J^{(\alpha)}(q,Y_\alpha)    \ .
\label{sumJ}
\end{equation}
The first $Y_\alpha$ incoherent sum rules are
\begin{eqnarray}
M_{0}^{(\alpha)} (q) &  =  & 1    
\label{m0J} \\
M_{1}^{(\alpha)} (q) &  =  & 0 
\label{m1J} \\
M_{2}^{(\alpha)} (q) & = & \frac{2m_\alpha}{3}
t_\alpha
\label{m2J} \\
M_{3}^{(\alpha)} (q) & = & \frac{m_\alpha \rho_\alpha}{2q}  \int d{\bf r} \, 
g^{(\alpha,\alpha)}(r) ({\bf q}
\cdot \nabla )^2 V(r)    \ . 
\label{m3J}
\end{eqnarray}
In the IA, $M_{0}^{(\alpha)}$, $M_{1}^{(\alpha)}$, and 
$M_{2}^{(\alpha)}$ coincide with the incoherent sum rules
(\ref{m0J},\ref{m1J},\ref{m2J}) but $M_{3,{\rm IA}}^{(\alpha)}=0$. The latter 
result
is in fact general for all the odd $Y_\alpha$-weighted sum rules in the 
IA due to the symmetry of the IA response around $Y_4=0$.

In a FSE convolutive theory, as the Gersch-Rodriguez one, it is easy to
extract the first sum rules of $R(q,Y)$. From the
total and the IA sum rules, the use of the algebraic relation
\begin{equation}
M_k(q) = \sum_{i=0}^{k}  {k \choose i}
 \, M_{i,{\rm IA}}(q) \, M_{k-i,{\rm R}} (q)
\label{combi}
\end{equation}
allows for the extraction of $M_{j,{\rm R}} (q)$:
\begin{eqnarray}
M_{0,{\rm R}} (q) &  =  & 1    
\label{m0R} \\
M_{1,{\rm R}} (q) &  =  & 0 
\label{m1R} \\
M_{2,{\rm R}} (q) & = & 0
\label{m2R} \\
M_{3,{\rm R}} (q) & = & \frac{m}{2q^3} \,  \rho \, \int d{\bf r} \, 
g(r) ({\bf q}
\cdot \nabla )^2 V(r)    \ . 
\label{m3R}
\end{eqnarray}
It can be proved that in the Gersch-Rodriguez prescription, 
the four moments (\ref{m0R},\ref{m1R},\ref{m2R},\ref{m3R}) are
exactly fulfilled.\cite{mazz1} It is worth noticing that $M_{3,{\rm R}} (q)$ is
satisfied if and only if a realistic two-body density matrix is used in the
calculation of $R(q,Y)$.

The theory proposed for $^3$He in the mixture (Sect. II) predicts a
response which is a sum of a convolution product plus a correction term
$\Delta S^{(3)}$. The function $R^{(3)}(q,Y_3)$ satisfies $M_{0,{\rm R}}(q)$, 
$M_{1,{\rm R}}(q)$, and $M_{2,{\rm R}}(q)$ but not $M_{3,{\rm R}}(q)$ because the
convolutive term relies on a boson-boson approximation. Concerning the
additive term $\Delta S^{(3)}$, it is straightforward to verify that their
three first moments are strictly zero whereas $M_3^{\Delta}(q)$ contains
corrections to the boson-boson $g^{(3,\alpha)}(r)$ functions assumed
in $M_{3,{\rm R}}(q)$.

\section{RESULTS}
The generalization of the Gersch-Rodriguez formalism to the $^3$He-$^4$He
mixture presented in Sect. II requires from the knowledge of microscopic
ground-state properties of the system. In the present work, the necessary input
has been obtained using the FHNC/HNC theory.\cite{polls,polls2} The variational wave
function is written as
\begin{equation}
\Psi = F \, \Phi_0    \ ,
\label{psi}
\end{equation}
with $F$ an operator that incorporates the dynamical correlations induced
by the interatomic potential, and $\Phi_0$ a model wave function that
introduces the right quantum statistics of each component. $\Phi_0$ is
considered a constant for bosons and a Slater determinant for fermions. In
the Jastrow approximation, the correlation factor $F$ is given by
\begin{equation}
F=F_{\rm J} = \prod_{\alpha \leq \beta} \prod_{i < j}
f_2^{(\alpha,\beta)}(r_{ij})    \ .
\label{jas}
\end{equation}
A significant improvement in the variational description of helium is
achieved when three-body correlations are included in the wave function.\cite{manu,usma} In this case,
\begin{equation}
F=F_{\rm JT} = \prod_{\alpha \leq \beta} \prod_{i < j}
f_2^{(\alpha,\beta)}(r_{ij}) \, \prod_{\alpha \leq \beta \leq \gamma} 
\prod_{i<j<k} f_3^{(\alpha,\beta,\gamma)}(r_{ij},r_{ik},r_{jk})   \ .
\label{jastri}
\end{equation}

The isotopic character of the mixture makes the interatomic
potential between the different pairs of particles be the same.
Therefore, the correlation factors $f_2^{(\alpha,\beta)}$ and 
$f_3^{(\alpha,\beta,\gamma)}$ can be considered  to first order as
independent of the indexes $\alpha$, $\beta$, $\gamma$. 
 That approach, known as average correlation approximation
(ACA),\cite{guyer} has been assumed throughout this work. DMC calculations of
$^3$He-$^4$He mixtures\cite{moro} have estimated that the influence of 
the ACA in the
momentum distributions is less than 5 \%.

The dynamic structure function of the mixture has been studied at $^3$He
concentrations $x=0.066$ and $x=0.095$ that, following the experimental
isobar\cite{ebner} $P=0$, correspond  to the total densities $\rho=0.3582\ \sigma^{-3}$ and
$\rho=0.3554\ \sigma^{-3}$ ($\sigma=2.556$ \AA), respectively. Notice the decrease of
$\rho$ when $x$ increases; in pure $^4$He, $\rho=0.3648\ \sigma^{-3}$. In
Table I, results for the $^4$He condensate fraction and kinetic energies
per particle are reported in J and JT approximations. The condensate
fraction increases with $x$ whereas the kinetic energies $t_\alpha$
decrease, both effects mainly due to the diminution of the
density. Results for pure  $^4$He  in the JT approximation (the one used
hereafter) compare favourably with DMC data from Ref. \onlinecite{boro3} ($n_0=0.084$,
$t_4=14.3$ K), and the decrease of $n_0$ with $x$ is in agreement with the
change in $n_0$ estimated using DMC.\cite{moro}

\subsection{Impulse Approximation}
One of the  characteristic properties of the IA in a pure system is its
$Y$-scaling. In this approximation, the response is usually written as the
Compton profile $J(Y)$. However,  global scaling
is lost in the mixture due to the different mass of the two helium isotopes.
 The
individual Compton profiles $J^{(\alpha)}(Y_\alpha)$ must be written in terms of
its own $Y_\alpha$ variable.

Results for $J^{(\alpha)}(Y_\alpha)$ at $x=0.095$ are shown in Fig. 1. The
different statistics of $^4$He and $^3$He are clearly visualized in their
respective momentum distributions, and therefore also in the Compton
profiles. In $J^{(4)}(Y_4)$, a delta singularity of
strength $n_0$ located at $Y_4=0$ (not shown in the figure)  emerges on top 
of the background, whereas
in $J^{(3)}(Y_3)$ the Fermi statistics is reflected in the kinks at
$Y_3=\pm k_F$ produced by the gap of $n^{(3)}(k)$ at $k=k_F$. The large
$|Y_\alpha|$ behavior of both responses is  more similar and is
entirely dominated by the tails of the respective momentum distributions.

The dynamic structure function of the mixture suggests the definition of a
total generalized Compton profiles $J(q,Y_\alpha)$.\cite{wang} In the IA,
\begin{equation}
J(q,Y_\alpha) = \frac{1}{\sigma_\alpha (\delta_{\alpha 3} x +
\delta_{\alpha 4} (1-x))} \, \frac{q}{m_\alpha} \, S_{\rm
IA}(q,\omega)  \ ,
\label{jiatot}
\end{equation}
with
\begin{equation}
S_{\rm IA}(q,\omega) = \sigma_4 (1-x) \, S_{\rm IA}^{(4)}(q,\omega)
+ \sigma_3  x \, S_{\rm IA}^{(3)}(q,\omega)  \ .
\label{siatot}
\end{equation}
Notice that the definition (\ref{jiatot}) is different for each $Y_\alpha$.
In order not to overload the notation, the introduction of a
new labelling in $J(q,Y_\alpha)$ has been omitted.
In terms of $Y_4$, and introducing the single Compton profiles
$J^{(\alpha)}(Y_\alpha)$,
\begin{equation}
 J(q,Y_4) = J^{(4)}(Y_4) + \frac{\sigma_3 x}{\sigma_4 (1-x)} 
 \, \frac{m_3}{m_4} \, J^{(3)}(Y_3(Y_4)) \ ,
\label{jtot4}
\end{equation}
with
\begin{equation}
Y_3(Y_4) = \frac{m_3}{m_4} Y_4 - \frac{q}{2} \left ( 1 - \frac{m_3}{m_4}
\right )         \ .
\label{y3y4}
\end{equation}
Equivalently, one can express the total generalized Compton profile as a function
of $Y_3$,
\begin{equation}
 J(q,Y_3) =  \frac{\sigma_4 (1-x)}{\sigma_3 x} \,  \frac{m_4}{m_3} 
    J^{(4)}(Y_4(Y_3)) +  J^{(3)}(Y_3) \ ,
\label{jtot3}
\end{equation}
with
\begin{equation}
Y_4(Y_3) = \frac{m_4}{m_3} Y_3 - \frac{q}{2} \left (  \frac{m_4}{m_3} - 1
\right )         \ .
\label{y4y3}
\end{equation}

The choice of the scaling variable $Y_\alpha$ undoubtedly determines some
trends of the response. If $Y_4$ is used, the $^4$He peak is centered at
$Y_4=0$ and the $^3$He peak shifts to $Y_4=(m_4/m_3 -1)\, q/2 \sim q/6$. On
the other side, if $Y_3$ is the choice the $^3$He peak is centered at
$Y_3=0$ and the $^4$He one moves to $Y_3=(m_3/m_4 -1)\, q/2 \sim -q/8$. In
addition, and disregarding cross sections and concentration factors, the
$^3$He peak is reduced by a factor $m_3/m_4$ when the response is expressed
in terms of $Y_4$. By the same token, the $^4$He peak is enhanced by a
factor $m_4/m_3$ when the response is written as a function of $Y_3$.

In Fig. 2, the IA responses for the mixture at two different $^3$He
concentrations are shown. They correspond to a momentum transfer $q=23.1$
\AA$^{-1}$ and have been obtained from $n^{(\alpha)}(k)$ calculated at the
JT approximation level. The differences between both curves are due to the
 concentration factors rather than to the differences between the
momentum distributions involved.

\subsection{Final State Effects}
The theory of FSE in $^3$He-$^4$He mixtures developed in Sect. II requires
from the knowledge of the three correcting functions $R^{(4)}(q,s)$ (\ref{rcon4}),
$R^{(3)}(q,s)$ (\ref{fse3}), and $\Delta S^{(3)}(q,s)$ (\ref{delta3l})
($s=t \, q/m_\alpha$). These three functions are complex with real and
imaginary parts that are, respectively, even and odd functions under the
change $s \rightarrow -s$. The latter is a consequence of the symmetry
properties of the two-body density matrices and of the central character of
the interatomic potential. The Fourier transforms of the real and imaginary
parts generate, respectively,  the even and odd components of
$R^{(\alpha)}(q,Y_\alpha)$ and $\Delta S^{(3})(q,Y_3)$, which are all
real.

In Fig. 3, the real and imaginary parts of $R^{(\alpha)}(q,s)$
corresponding to a $x=0.095$ mixture are shown. In spite of the fact that 
$R^{(4)}(q,s)$ is calculated for the real mixture and $R^{(3)}(q,s)$ 
for the boson-boson one, the differences between the two functions are
rather small. Actually, those differences are mainly attributable to the low $^3$He
density in the mixture that makes the contributions of the Fermi statistics
very small.
In fact, the differences shown in Fig. 3 between $R^{(4)}(q,s)$ and 
$R^{(3)}(q,s)$ are essentially due to the different mass of the two
isotopes, which  factorizes in the integral of the interatomic potentials
(see Eqs.\ref{rcon4},\ref{fse3}).

The real and imaginary parts of the additive term $\Delta S^{(3)}(q,s)$ are
shown in Fig. 4 at the two $^3$He concentrations studied. The behavior of 
$\Delta S^{(3)}(q,s)$ is remarkably different from the behavior of the FSE
broadening functions $R^{(\alpha)}(q,s)$, presenting oscillating tails that
slowly fall to zero with increasing $x$. The function $\Delta S^{(3)}(q,s)$
incorporates on the $^3$He response all the Fermi corrections which are not
contained in $R^{(3)}(q,s)$. In a dilute Fermi liquid, as $^3$He in the
mixture, those contributions are characterized by the behavior of $l(k_F
r)$ and $l^2(k_F r)$, $l(z)=3/\,z^3\, (\sin z - z \cos z)$ being the Slater
function.

$R^{(4)}(q,Y_4)$ and $R^{(3)}(q,Y_3)$ are compared in Fig. 5 at $x=0.095$ and
$q=23.1$ \AA$^{-1}$. The shape of both functions looks very  much the same: a
dominant central peak and small oscillating tails that vanish with
$|Y_\alpha|$. The figure also shows that at a given concentration the central peak of
$R^{(3)}(q,Y_3)$ is slightly higher and narrower than the one of 
$R^{(4)}(q,Y_4)$, an effect once again due to the different mass of 
the two isotopes. Therefore, at a
fixed momentum transfer $q$ FSE in $^4$He are expected to  be
smaller than in $^4$He. In the scale used in Fig. 5, the
$R^{(\alpha)}(q,Y_\alpha)$ functions at $x=0.066$ would be hardly
distinguishable from the ones at $x=0.095$.

The Compton profile $\Delta J^{(3)}(q,Y_3)$, derived from the Fourier
transform of $\Delta S^{(3)}(q,s)$, is shown in Fig. 6 at the two $x$ values
considered. $\Delta J^{(3)}(q,Y_3)$ presents a central peak and two minima
close to $Y_3=\pm k_F$. The absolute value of this function is small
compared to both $R^{(3)}(q,Y_3)$ and the IA response $J^{(3)}(Y_3)$ (\ref{jia}) 
but manifests a sizeable dependence on the $^3$He concentration. This
feature is patent in Fig. 6, where one can see how the contribution of 
$\Delta J^{(3)}(q,Y_3)$ increases with $x$. This is an expected result
taking into account that in the current approximation $\Delta
J^{(3)}(q,Y_3)$ incorporates all the Fermi effects to the $^3$He FSE
function.
 
According to the theory developed in Sect. II, the $^4$He response in the
mixture, $J^{(4)}(q,Y_4)$ is the sum of two terms: the non-condensate part
of the IA convoluted with $R^{(4)}(q,Y_4)$, and $n_0 R^{(4)}(q,Y_4)$, which
is the contribution of the condensate once broadened by FSE. 
The different terms contributing to the final response are separately shown
in Fig. 7.
The
correction driven by $n_0$ is by far the largest one. In spite of the small
value of $n_0$, the broadening of the condensate term, which transforms the
delta singularity predicted by the IA into a function of finite height
and width, unambiguously produces non-negligible FSE in the $^4$He peak.

The obvious lack of a condensate fraction  in the $^3$He component reduces
the quantitative relevance of its FSE. The $^3$He FSE correcting functions
and the corresponding IA response, are compared in Fig. 8 at $x=0.095$. The
convolution of the IA with $R^{(3)}(q,Y_3)$ produces a slight quenching of
$J^{(3)}(q,Y_3)$ around the peak and a complete smoothing of the
discontinuity in the derivative of $J^{(3)}(Y_3)$ at $Y_3=\pm k_F$. The
contribution of $\Delta J^{(3)}(q,Y_3)$ is rather small but
restores to some extent the change in the derivative around $k_F$.

\subsection{Theory vs. Experiment}
Scattering experiments suffer from instrumental resolution effects (IRE)
that tend to smooth the detailed structure of the dynamic structure
function. Any comparison between theory and experiment have therefore 
to include in
the analysis the IRE contributions. From the theoretical side, it would be
desirable to remove the IRE from the data to allow for a direct comparison.
This process would imply a deconvolution procedure that is known to be highly
unstable. As suggested by Sokol {\em et al.},\cite{ire} it is better to convolute
the theoretical prediction with the IRE function
$I^{(\alpha)}(q,Y_\alpha)$, and then to compare the result with the
experimental data. The functions $I^{(\alpha)}(q,Y_\alpha)$
provided by Sokol\cite{priv} 
 are reported in Fig. 9. As one can see, at $q=23.1$
\AA$^{-1}$ the IRE corrections are of the same order of the FSE functions
$R^{(\alpha)}(q,Y_\alpha)$, and in fact their magnitude significantly
increases with $q$. The IRE functions for the mixture (Fig. 9) present a
small shift of their maximum to negative $Y$ values, a feature that makes the
peak of the total response be slightly moved in the same direction.

In Fig. 10, the generalized Compton profile $J(q,Y_4)$ (including both the
IRE and FSE) is compared with the scattering data of Wang and
Sokol.\cite{wang} Those
measurements were carried out in a $x=0.095$ mixture at $T=1.4$ K and 
a momentum transfer $q=23.1$ \AA$^{-1}$. The analysis of the experimental
data led the authors to estimate the $^4$He condensate fraction and the
single-particle kinetic energies of both species. In Ref.
\onlinecite{wang},
 a value
$n_0=0.18 \pm 0.03$, and kinetic energies $t_4=13 \pm 3$ K and $t_3=11 \pm
3$ K are reported. That work, and an independent measurement performed by
Azuah {\em et al.},\cite{stirli} agree in the values of the kinetic energies and in
their dependence with the $^3$He concentration. Both analysis coincide in a
decrease in $t_4$ with $x$ and a more surprising constancy of $t_3$ along
$x$. Microscopic calculations\cite{boro3} of those quantities only agree with the
experimental result of $t_4(\rho)$. Several independent
calculations,\cite{boro2,moro}
including the present one, suggest smaller values of $n_0$ ($n_0 \simeq
0.10$) and larger values of $t_3$ ($t_3 \simeq 18$ K), in clear
disagreement with the experimental estimations.

Let us turn to Fig. 10 with the comparison between the theoretical and
experimental responses. The theoretical result, constructed using Eqs.
(\ref{jtot4},\ref{jtot3}), but replacing the IA $J^{(\alpha)}(Y_\alpha)$
with the final responses $J^{(\alpha)}(q,Y_\alpha)$, shows sizeable
differences with respect to the experimental data and a lack of strength
 below the two peaks. In
order to clarify the origin of such a large discrepancy, we have compared
the $M_0$ and $M_1$ sum rules obtained by direct integration of the
experimental $J(q,Y_4)$ with the theoretical results (Sect. III). That
check has shown that the $M_0$ and $M_1$ values obtained from the two
procedures are not compatible. Our conclusion is that the reported
experimental Compton profiles are probably written in in a different
way than in Eq. (\ref{jiatot}). In fact, after the analysis of different 
possibilities, we have verified that
if one defines the response in the form
\begin{equation}
\tilde{J}(q,Y_4) = J^{(4)}(q,Y_4) + \frac{\sigma_3 x}{\sigma_4 (1-x)} \, 
J^{(3)}(q,Y_3(Y_4))
\label{jtilde4}
\end{equation}
or
\begin{equation}
\tilde{J}(q,Y_3) =  \frac{\sigma_4 (1-x)}{\sigma_3 x}\, J^{(4)}(q,Y_4(Y_3)) + 
J^{(3)}(q,Y_3)     \ ,
\label{jtilde3}
\end{equation}
the agreement in both sum rules is recovered. By moving our results to those
modified Compton profiles $\tilde{J}(q,Y_\alpha)$, the agreement between
theory and experiment improves significantly but only to what concerns the
$^3$He peak. Notice that the $^4$He peak is not modified when going from 
$J(q,Y_4)$ to $\tilde{J}(q,Y_4)$, and that a significant difference in the
height of the peak still remains.

The missing strength of the theoretical $^4$He peak with respect to the
experimental data could justify the difference between the theoretical and
experimental values of $n_0$. However, the present variational momentum
distribution predicts  $n_0$ values that are indistinguishable from a DMC
estimation.\cite{moro} Therefore, this difference should  not be attributed
to inaccuracies of our
$n^{(\alpha)}(k)$ but rather to an intriguing gap between theory and experiment.
At this point, it is worth considering the difficulties the experimentalists
have to face to extract $n_0$ and $t_\alpha$ from the measured data.
On the one hand,  experience in the pure $^4$He response has
shown that different momentum distributions (with different $n_0$'s) can be
accurately fitted to the data. On the other, the kinetic energy per
particle is derived from the $Y_\alpha^2$ sum rule whose estimation  is
highly influenced by the tails of the response. Those tails cannot be
accurately resolved due to the noise of the data, and thus the prediction
of $t_\alpha$ appears relatively uncertain. That is even more pronounced in
the $^3$He peak because the strong interaction with $^4$He causes
$n^{(3)}(k)$ to present non-negligible occupations up to large $k$ values.

The influence of $n_0$ and $t_\alpha$ on the momentum distribution, and
hence on the response, can be roughly estimated from the behavior of the
one-body density matrix. In a simple approximation, one can perform a
cumulant expansion of $\rho_1^{(\alpha)}(r)$ and relate the lowest order
cumulants to the lowest order sum rules of $n^{(\alpha)}(k)$. 
Introducing an expansion parameter $\lambda$,
\begin{equation}
\frac{1}{\rho_4}\,\rho_1^{(4)}(\lambda r) - n_0
\equiv e^{\mu_0 + \lambda^2\mu_2 + \cdots} =
\left( 1-n_0 \right) - \lambda^2 \left\langle \left( 
{\bf r} \cdot {\bf p}_1 \right)^2 \right\rangle + \cdots 
  \ .
\label{rtil1}
\end{equation}
 Taking into account that
\begin{equation}
\left\langle ({\bf r} \cdot {\bf p}_1)^2 \right\rangle = 
\frac{2m_4 r^2}{3}  t_4  \ ,
\label{rtil3}
\end{equation}
and considering $\lambda=1$,
\begin{equation}
\frac{1}{\rho_4}\,\rho_1^{(4)}(r) = n_0 + (1-n_0) 
\exp \left[ -\frac{2}{3}\frac{m_4 r^2}{(1-n_0)}
 t_4  + \cdots
\right] \ .
\label{rtil4}
\end{equation}

Equation (\ref{rtil4}) can then be used to relate $\rho_1^{(4)}(r)$ to a
new one-body density matrix $\overline{\rho}_1^{(4)}(r)$ with slightly
different values $\overline{n}_0$ and $\overline{t}_4$
\begin{equation}
\frac{1}{\rho_4}\,\overline{\rho}_1^{(4)}(r) = \overline{n}_0 + 
\left(\frac{1}{\rho_4}\,\rho_1^{(4)}(r) - n_0 \right)
\left(\frac{1-\overline{n}_0}{1-n_0}\right) \exp\left[
-\frac{2m_4 r^2}{3}
\left(\frac{\overline{t}_4}{1-\overline{n}_0}
- \frac{ t_4 }{1-n_0} \right) + \cdots \right]  \ .
\label{rtil5}
\end{equation}
In this way, the perturbed $\overline{\rho}_1^{(4)}(r)$ and
$\overline{n}^{(4)}(k)$ preserve their normalization and allows one to go
beyond a simple $n_0$ re-scaling. Using this method, we
have studied the effect of changing $n_0$ and $t_4$ on the $^4$He response.
 In Fig. 11, the results corresponding to {\it i})
$n_0=0.14$, $t_4=13.9$ K, and {\it ii}) $n_0=0.10$, $t_4=13.0$ K are shown. As one
can see, both slight changes in the theoretical response lead to a nice
agreement with the experimental data. Consequently, such a large value of
$n_0$ ($n_0^{\rm expt}=0.18$) does not seem to be required in order to
reproduce the additional strength observed below the $^4$He peak. The
re-scaling (\ref{rtil5}) shows that a small decrease in the kinetic
energy enhances the central peak in the same form  an increase of the
condensate fraction does.

\section{SUMMARY AND CONCLUSIONS}
A generalized Gersch-Rodriguez formalism has been applied to study the
dynamic structure function of the $^3$He-$^4$He mixture at high momentum
transfer. The Fermi character of $^3$He forbids a  straightforward
generalization of most  FSE theories used in bosonic systems, a problem that has 
been overcome in an approximate way. The
approximations assumed are however expected to include the leading Fermi
contributions to the FSE, at least in the mixture where the $^3$He partial
density is very small.

The theoretical response obtained shows significant differences with
scattering data in both the $^4$He and the $^3$He peaks. However, a sum-rules
analysis of the experimental response has shown some inconsistencies.
Redefining the total response,  it is possible to reach agreement between
the theoretical and
the numerical values of the first-order sum rules. If the theoretical 
response is
changed in the same way, the agreement is much better.
Nevertheless, the $^4$He peak is not modified by this redefinition (written
as a function of $Y_4$) and an intriguing sizeable difference in its
strength subsists. From the theoretical side, several arguments may be
argued trying to explain the observed discrepancies. The first uncertainty
could be attributed to the use of a Gersch-Rodriguez theory to account for
the FSE. In our opinion, that criticism has probably no sense because we
have verified that, at similar momentum transfer, the experimental response
of pure $^4$He is fully recovered with the GR theory.\cite{mazz1} Assuming therefore that
the theoretical framework is able to describe the high-$q$ response of the
mixture, one could be led to argue that the approximate microscopic inputs
of the theory are not accurate enough. That argument was put forward in
Ref.
\onlinecite{wang} to explain the differences in $n_0$ and $t_3$. One of the main
criticisms was the use of the ACA, which they claimed could be too
restrictive to allow for a reduction of $t_3$ towards a value closer to the
experimental one. However, a DMC calculation\cite{imp3} in which the ACA is 
 not
present, has proved that only a diminution of $\sim 0.5$ K in $t_3$ is obtained. 
Concerning the condensate fraction value,  our variational theory predicts a slight
increase of $n_0$ with $x$. This increase, which is mainly due to the
decrease of the equilibrium density when $x$ grows, is nevertheless much
 smaller then the one that would be required to reproduce the experimental
prediction. Our results for $n_0$ are again in an overall agreement with
the nearly exact DMC calculation of Ref. \onlinecite{moro}. 

In summary, we would like to emphasize
that there exists theoretical agreement on the values of $n_0$ and
$t_3$ for mixtures, but these values are quite far from the experimental
estimations. Additional scattering measurements on the $^3$He-$^4$He mixture
are necessary to solve the puzzle.

\acknowledgments
This research has been partially supported by DGESIC (Spain) Grants N$^0$
 PB98-0922 and PB98-1247, and DGR (Catalunya) Grants N$^0$
1999SGR-00146 and SGR99-0011. F. M. acknowledges the support from the
Austrian Science Fund under Grant N$^0$ P12832-TPH.

\begin{table}

\caption{Condensate fraction and kinetic energies as a function of $x$. At
each $^3$He concentration $x$ the first row corresponds to the J
approximation and the second one to the JT one.}

\begin{tabular}{lcccc}
 $x$ & $\rho(\sigma^{-3})$ & $n_0$ & $t_3$ (K) & $t_4$ (K)  \\  \tableline
0       &  0.3648     &  0.091   &       &  15.0     \\
        &             &  0.082   &      &  14.5     \\ \tableline
0.066    & 0.3582      & 0.095    &   19.9    &   14.6   \\
         &             & 0.088    &  18.7    &   14.1    \\ \tableline 
0.095    &  0.3554   &   0.097     &   19.6   &   18.5  \\
         &           &   0.090     &   18.5    &  13.9   
\end{tabular}

\end{table}

\begin{figure}
\caption{Compton profiles of $^4$He (left) and $^3$He (right), both in JT
(solid line) and J (dashed line) approximations  for $x$=0.095.}
\begin{center}
\epsfxsize=16cm  \epsfbox{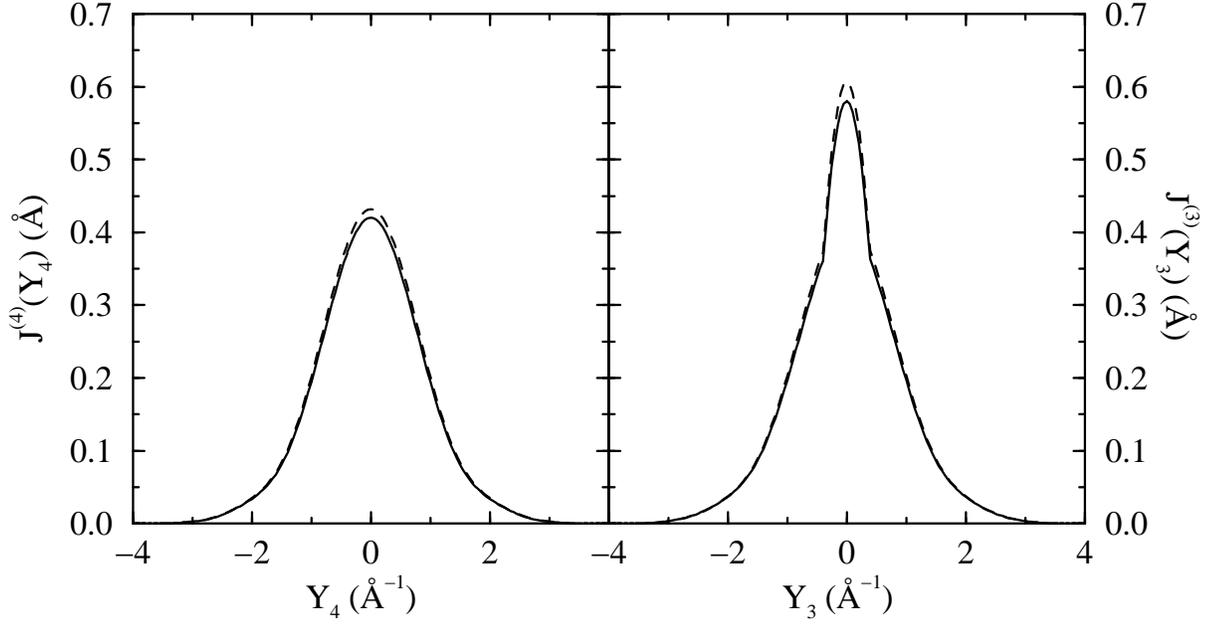}
\end{center}                                      
\end{figure}

\begin{figure}
\caption{Generalized Compton profiles in IA at $x=0.066$ (left) and $x=0.095$
(right).}
\begin{center}
\epsfxsize=16cm  \epsfbox{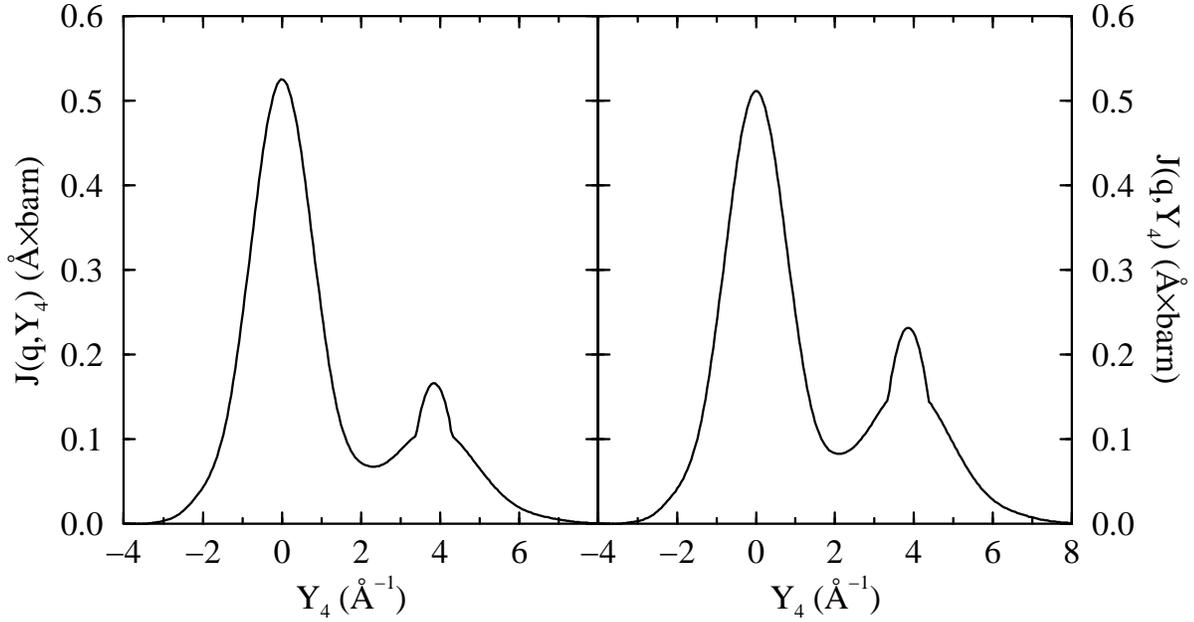}
\end{center}                                      
\end{figure} 

\pagebreak

\begin{figure}    
\caption{Real and Imaginary parts of $R^{(3)}(q,s)$ (solid line)
    and $R^{(4)}(q,s)$ (dashed line) at $q = 23.1$ \AA$^{-1}$ 
    for the $x = 0.095$ mixture.}
\begin{center}
\epsfxsize=16cm  \epsfbox{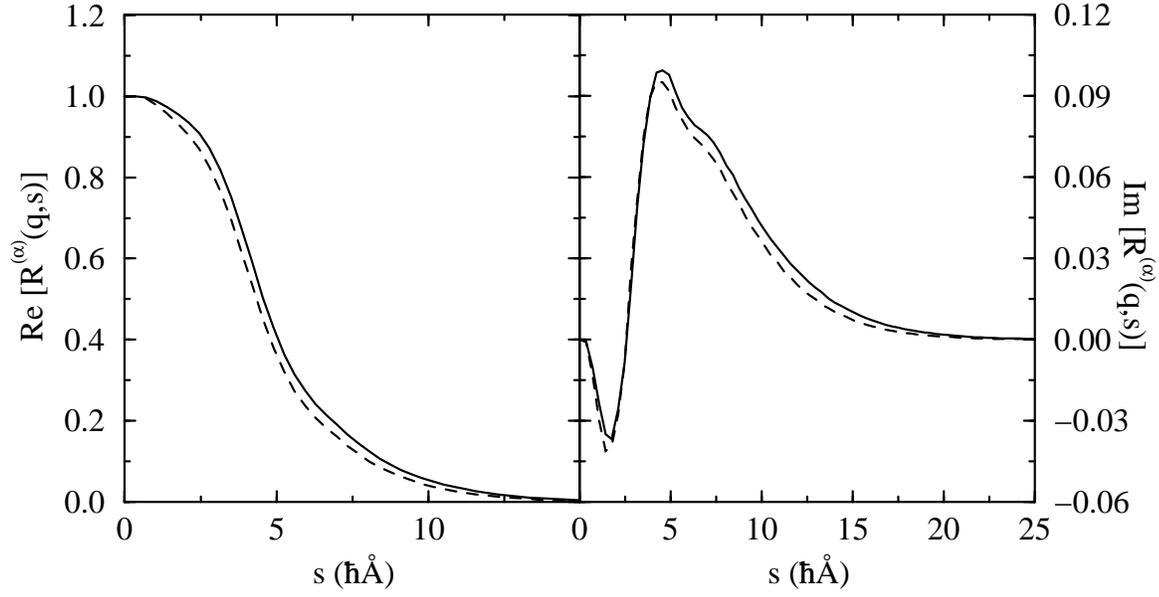}
 \end{center}
\end{figure}

\begin{figure}
 \caption{$\Delta S^{(3)}(q,s)$ at $q=23.1$ \AA$^{-1}$ and
    for  mixtures at $x = 0.095$ and $x = 0.066$ (solid
    and dashed lines).} 
 \begin{center}
 \epsfxsize=16cm  \epsfbox{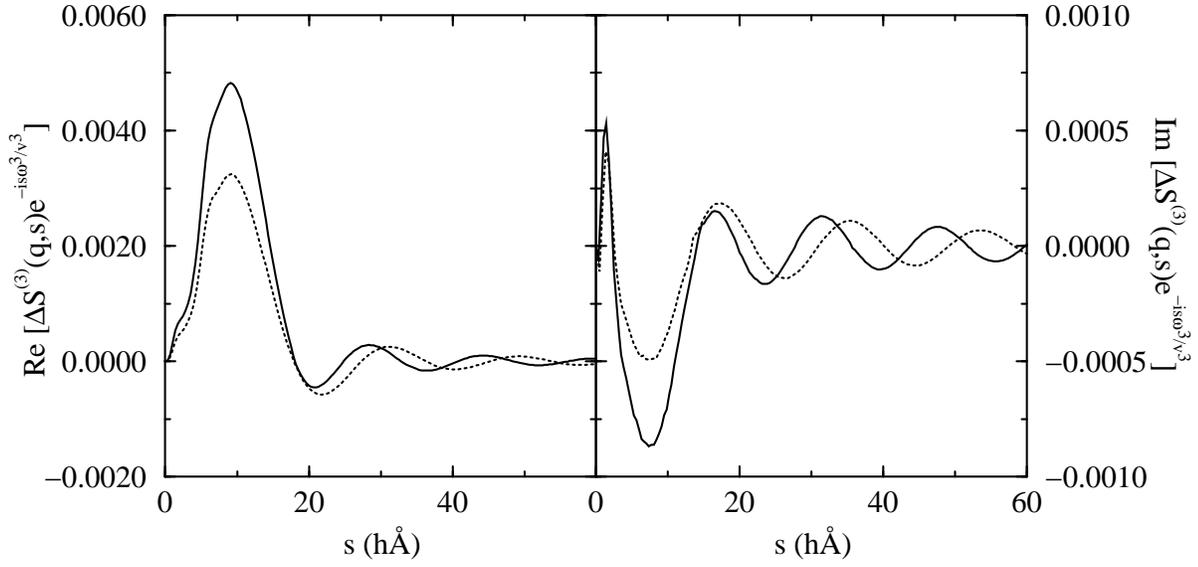}
   \end{center}
\end{figure}

\pagebreak

\begin{figure}
 \caption{Comparison between $R^{(4)}(q,Y_4)$ and
    $R^{(3)}(q,Y_3)$ at $q = 23.1$ \AA$^{-1}$ and for 
    $x = 0.095$ (solid and dashed lines,
    respectively). Notice that different $Y_\alpha$ variables 
     are used to depict each function.} 
  \begin{center}
 \epsfxsize=10cm  \epsfbox{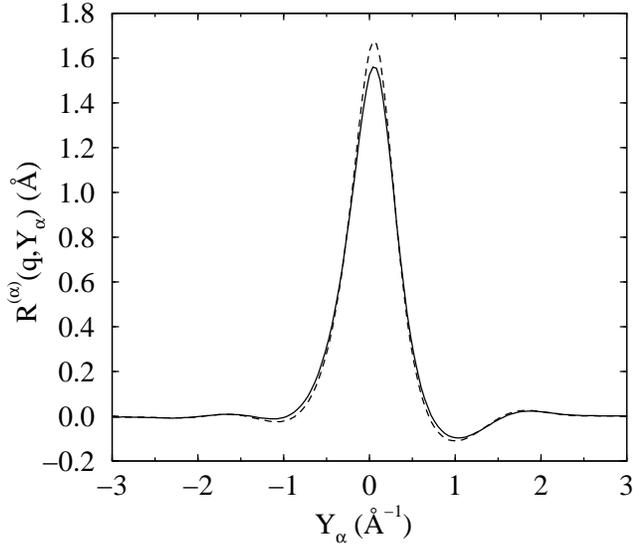} 
  \end{center}
\end{figure}


\begin{figure}
 \caption{The $^3$He additive correcting term 
    at $q = 23.1$ \AA$^{-1}$ for  $x = 0.095$ and 
    $x = 0.066$ mixtures (solid and dashed lines, respectively).} 
  \begin{center}
 \epsfxsize=10cm  \epsfbox{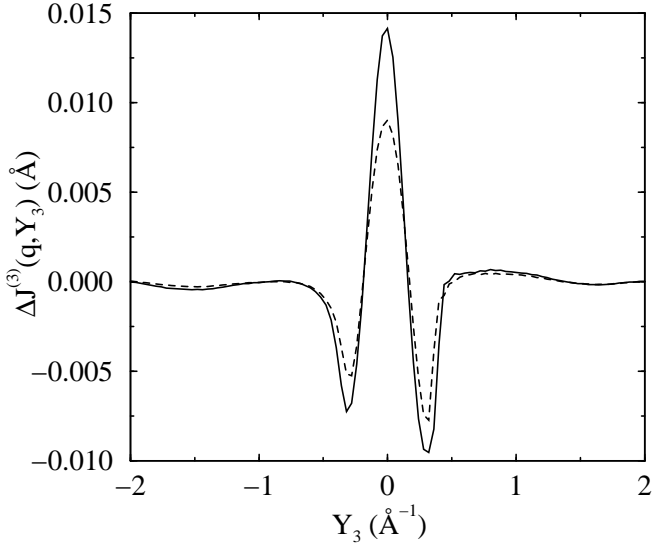}  
  \end{center}
\end{figure}

\pagebreak

\begin{figure}
  \caption{The different contributions to the $^4$He response at
     $x = 0.095$ and $q = 23.1$ \AA$^{-1}$. Dotted
    line: $^4$He Compton profile, dashed line: the same convoluted
    with $R^{(4)}(q,Y_4)$, long-dashed line: $n_0 R^{(4)}(q,Y_4)$,
    solid line: total $^4$He response.} 
  \begin{center}
   \epsfxsize=10cm  \epsfbox{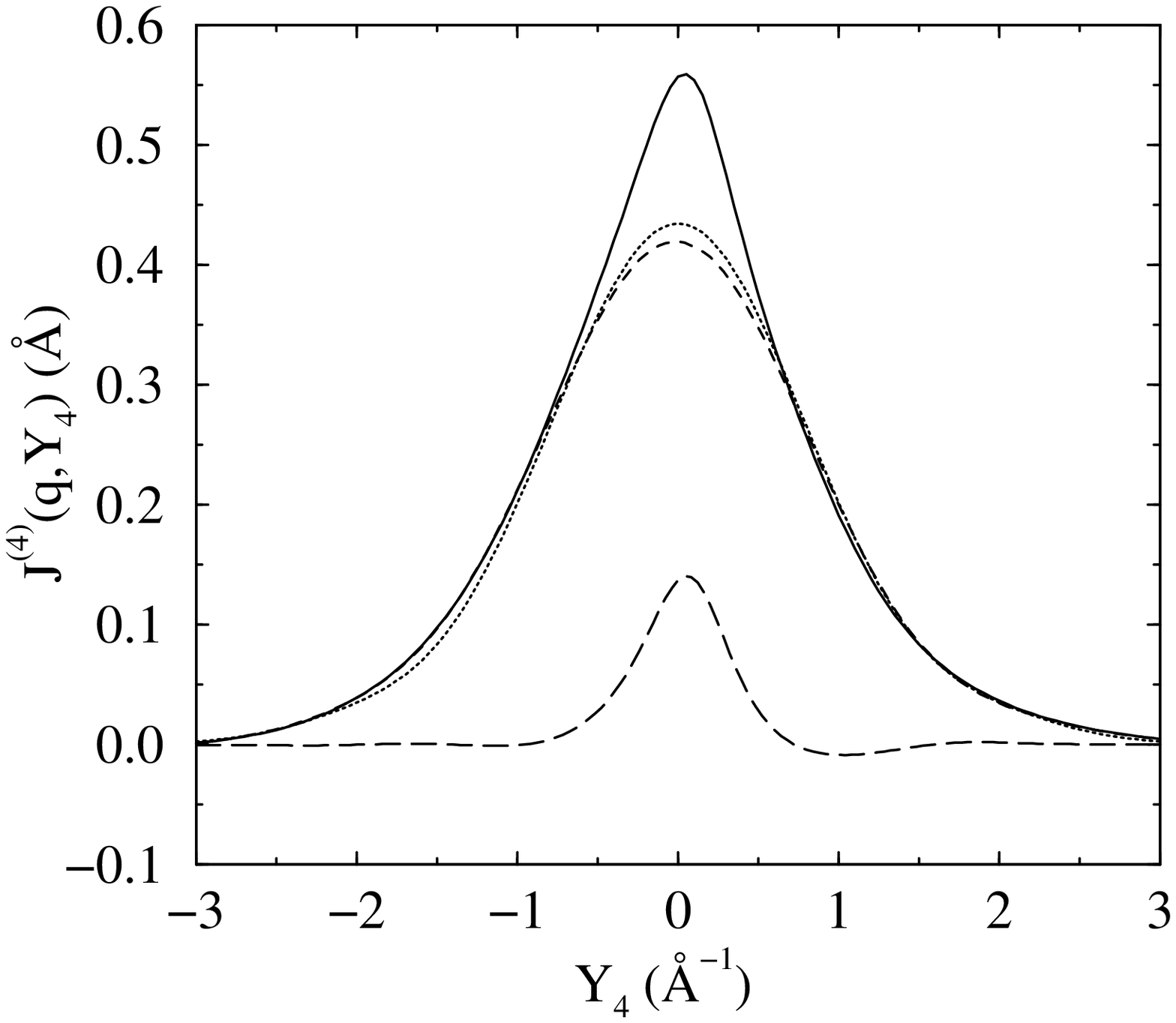}   
   \end{center}
\end{figure}


\begin{figure}
  \caption{The different contributions to the $^3$He response at
    $x = 0.095$ and  $q = 23.1$ \AA$^{-1}$. Dotted
    line: $^3$He Compton profile, dashed line: the same convoluted
    with $R^{(3)}(q,Y_4)$, dotted-dashed line: $\Delta
    J^{(3)}(q,Y_3)$, solid line: total $^3$He response.} 
  \begin{center} 
  \epsfxsize=10cm  \epsfbox{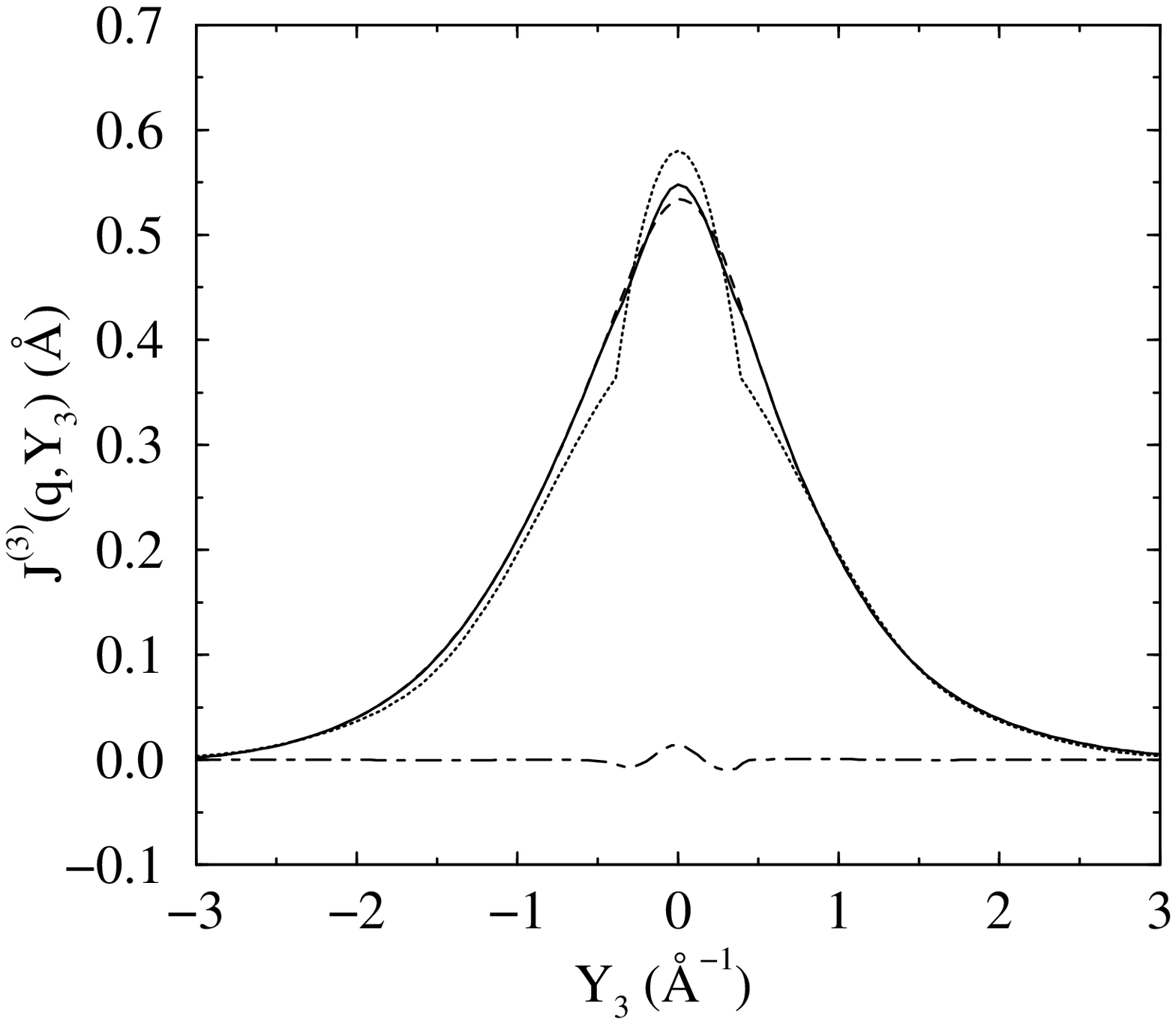}    
  \end{center}
\end{figure}

 \pagebreak

\begin{figure}
  \caption{$^4$He (left) and $^3$He (right) instrumental resolution
  functions at $x=0.095$ and $q=23.1$ \AA$^{-1}$.} 
  \begin{center} 
  \epsfxsize=16cm  \epsfbox{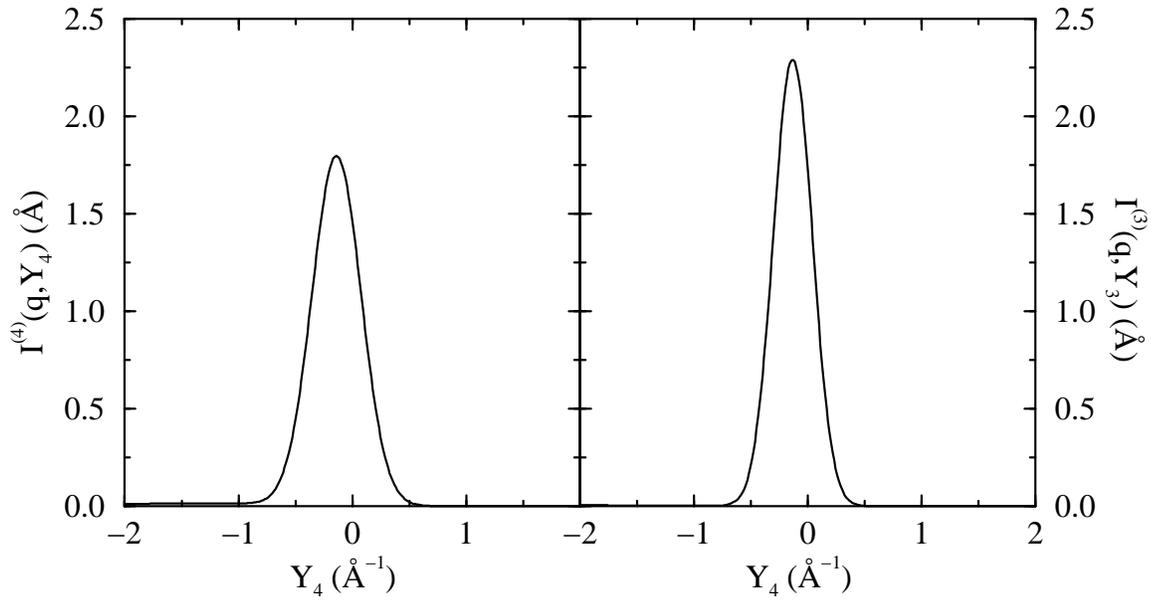}
  \end{center}
\end{figure}


\begin{figure}
  \caption{Comparison of the theoretical generalized Compton
    profile (solid line) and the experimental measurements of Wang and
    Sokol\protect\cite{wang} of the $x = 0.095$ mixture  at
    $q = 23.1$ \AA$^{-1}$ and $T = 1.4$K (points with errorbars).} 
    \begin{center}
   \epsfxsize=10cm  \epsfbox{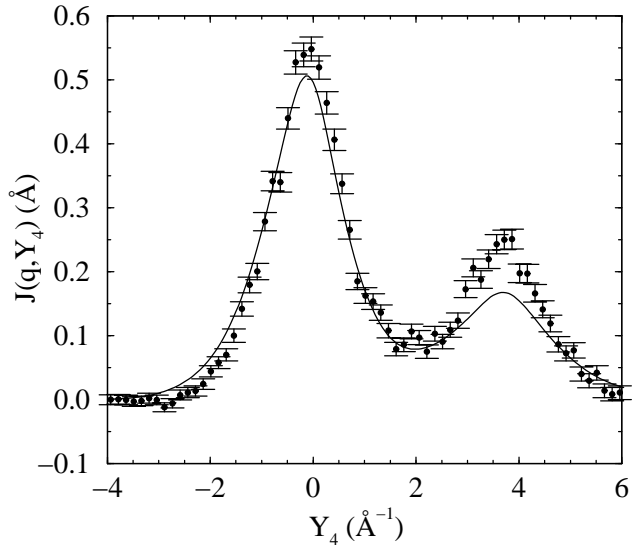}   
   \end{center}
\end{figure}

\pagebreak

\begin{figure}
 \caption{The $x = 0.095$ experimental data of Wang and
 Sokol\protect\cite{wang}
    compared to the response obtained from an alternative
    $\overline{\rho}_1^{(4)}(r)$ with $n_0 = 0.14$ and
    $\overline{t}_4=  t_4= 13.9$ K (solid line) --left panel--,
     and with $n_0 = 0.10$ and
    $\overline{t}_4=  13.0$ K (solid line) --right panel--.} 
  \begin{center}
  \epsfxsize=16cm  \epsfbox{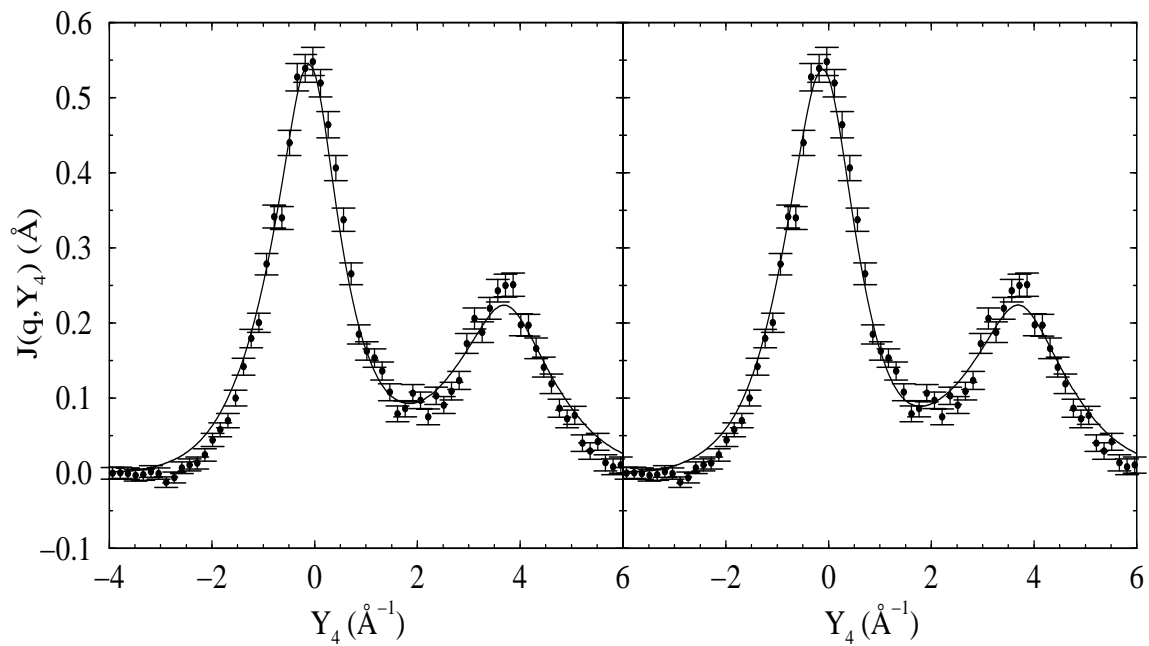}
    \end{center}
\end{figure}

\end{document}